\documentclass{vgtc}                          %

\ifpdf%
  \pdfoutput=1\relax                   %
  \usepackage{graphicx}                %
  \DeclareGraphicsExtensions{.pdf,.png,.jpg,.jpeg} %
\else%
  \usepackage{graphicx}                %
  \DeclareGraphicsExtensions{.eps}     %
\fi%

\graphicspath{{figures/}{pictures/}{images/}{./}} %

\usepackage{balance}
\usepackage{microtype}                 %
\PassOptionsToPackage{warn}{textcomp}  %
\usepackage{textcomp}                  %
\usepackage{mathptmx}                  %
\usepackage{times}                     %
\usepackage{cite}                      %
\usepackage{tabu}                      %
\usepackage{booktabs}                  %

\onlineid{0}

\vgtccategory{Research}

\vgtcinsertpkg
 \usepackage{float}

\title{Does Adding Physical Realism to Virtual Reality\\Training Reduce Time Compression?}

\author{Kadir Lofca\thanks{e-mail: k\_lofca@uncg.edu}\\ %
        \scriptsize Dept. of Computer Science\\\scriptsize UNC Greensboro %
     \and Jason Jerald\thanks{e-mail: jason@nextgeninteractions.com}\\ %
     \scriptsize NextGen Interactions %
     \and Dalton Costa\thanks{e-mail: dbcosta@uncg.edu}\\ %
     \scriptsize Dept. of Computer Science\\\scriptsize UNC Greensboro
     \and Regis Kopper\thanks{e-mail: kopper@uncg.edu}\\ %
     \scriptsize Dept. of Computer Science\\\scriptsize UNC Greensboro %
     }

\teaser{
 \includegraphics[height=1.85in]{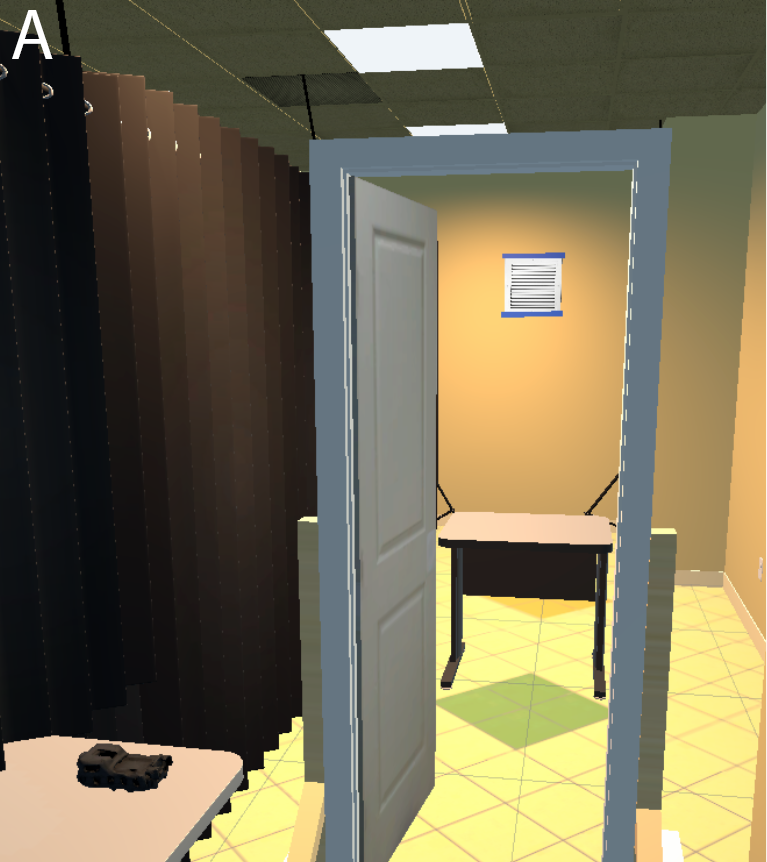}~~
 \includegraphics[height=1.85in]{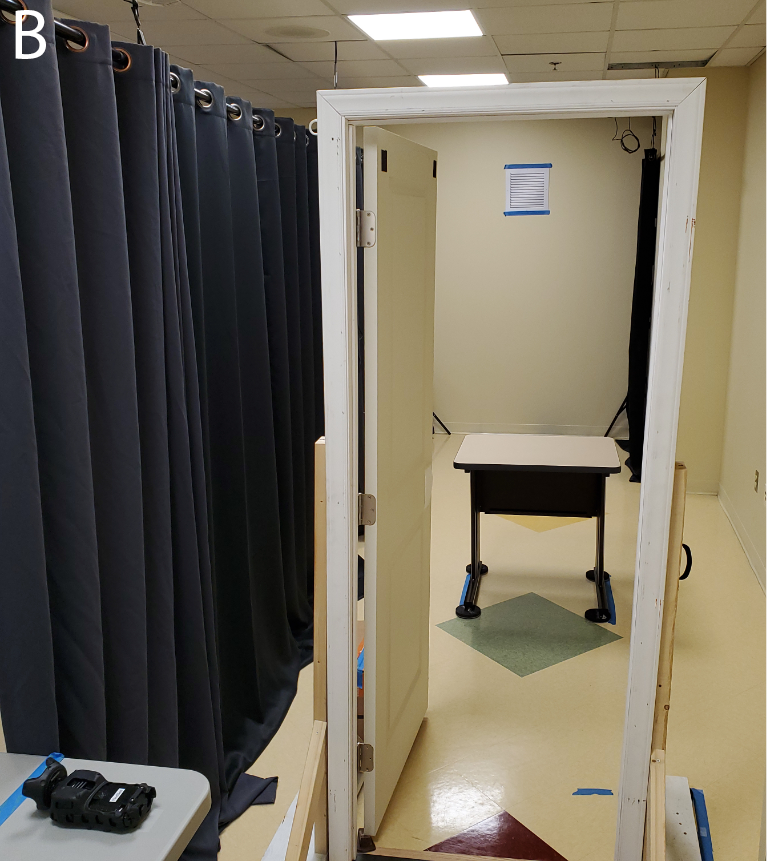}~~
 \includegraphics[height=1.85in]{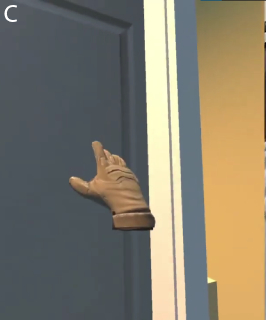}~~
 \includegraphics[height=1.85in]{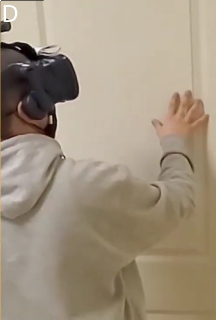}
 \caption{Hazmat training virtual and real environments. The virtual environment was modeled as a precise copy of the real one. In the high physical realism VR condition, the user picks up a real gas monitor prop (B, lower left.) which is tracked and registered to the virtual counterpart (A, lower left). The tracked hands are mapped to virtual hands, and the user pushes a real tracked door (D) which animates its virtual counterpart (C).}
 \label{fig:RealMultiRAE}
}

\abstract{
Virtual reality (VR) is known to cause a ``time compression'' effect, where the time spent in VR feels to pass faster than the effective elapsed time. Our goal with this research is to investigate if the physical realism of a VR experience reduces the time compression effect on a gas monitoring training task that requires precise time estimation. We used physical props and passive haptics in a VR task with high physical realism and compared it to an equivalent standard VR task with only virtual objects. We also used an identical real-world task as a baseline time estimation task. Each scenario includes the user picking up a device, opening a door, navigating a corridor with obstacles, performing five short time estimations, and estimating the total time from task start to end. Contrary to previous work, there was a consistent time dilation effect in all conditions, including the real world. However, no significant effects were found comparing the estimated differences between the high and low physical realism conditions. We discuss implications of the results and limitations of the study and propose future work that may better address this important question for virtual reality training.%
} %

\CCScatlist{
  \CCScatTwelve{Human-centered computing}{Time-perception}{Virtual Reality}{};
  \CCScatTwelve{Human-centered computing}{Human computer interaction (HCI)}{Interaction devices}{Haptic devices}.
}

\begin{document}

\firstsection{Introduction}

\maketitle

Hazardous material (HAZMAT) technicians respond to emergencies where their perception of time is essential to the success of the operation. One example scenario is the identification of gas leaks. In order to correctly identify a potential gas leak, HAZMAT technicians must hold a monitoring tool close to a potential source for a sufficient amount of time.  Failing to hold the monitoring tool long enough could lead to undetected leaks that can have severe and potentially deadly consequences. At the same time, emergency calls that result in severe gas leaks are not common\footnote{The average number of methane gas leaks in the US per year between 2010 and 2020 was approximately 230~\cite{Dutzik:2022aa}, which is high for its environmental and safety impacts, but rare enough that most first responders rarely respond to emergency calls involving leaks}, which causes HAZMAT technicians to gain little practical experience real emergency situations. Traditional training using classroom instruction and real-life simulation exercises may not be as effective as VR-based training, where the user can have real-time visual feedback about a potentially missed leak occurrence.

However, VR causes a well-known \emph{time compression} effect, where the perception of time by a user in VR is perceived as shorter than the actual elapsed time~\cite{schneider2011effect, mullen2021time}. In other words, much like watching a movie or playing a video game, VR can make time feel as passing faster for the user. So, in situations where measuring time is essential to the learning of a task, as is the case with gas monitoring, such a time compression effect could lead to negative results. For example, HAZMAT technician trainees might learn in VR that what they perceive as being a sufficient amount of time to detect a gas leak (say, 30s) is actually a shorter amount of time (say, 25s). This could lead to a failed execution of the task in a real-life situation. 

Thus, our goal with this study was to quantify time estimation under different conditions in VR, as compared to a real-world equivalent task~\cite{lofca2022studying}.
Next, we discuss relevant VR literature on training, physical props and passive haptics, and time perception and estimation.

\subsection{VR Training}
 Different from other forms of computer interfaces, VR translates detailed human body motion directly into the virtual environment and creates a highly realistic experience. Coupled with the ever-increasing quality of computer graphics, VR is capable of simulating life-like experiences. Virtual reality-based training is getting increasingly popular in fields where re-creating specific scenarios is costly, risky, or inconvenient, such as in minimally invasive medical procedures~\cite{alvarez2020use}, the military~\cite{rettinger2022defuse}, and public safety~\cite{grandi2021design}.
 
\begin{figure}[htb]
 \centering%
 \includegraphics[width=\columnwidth]{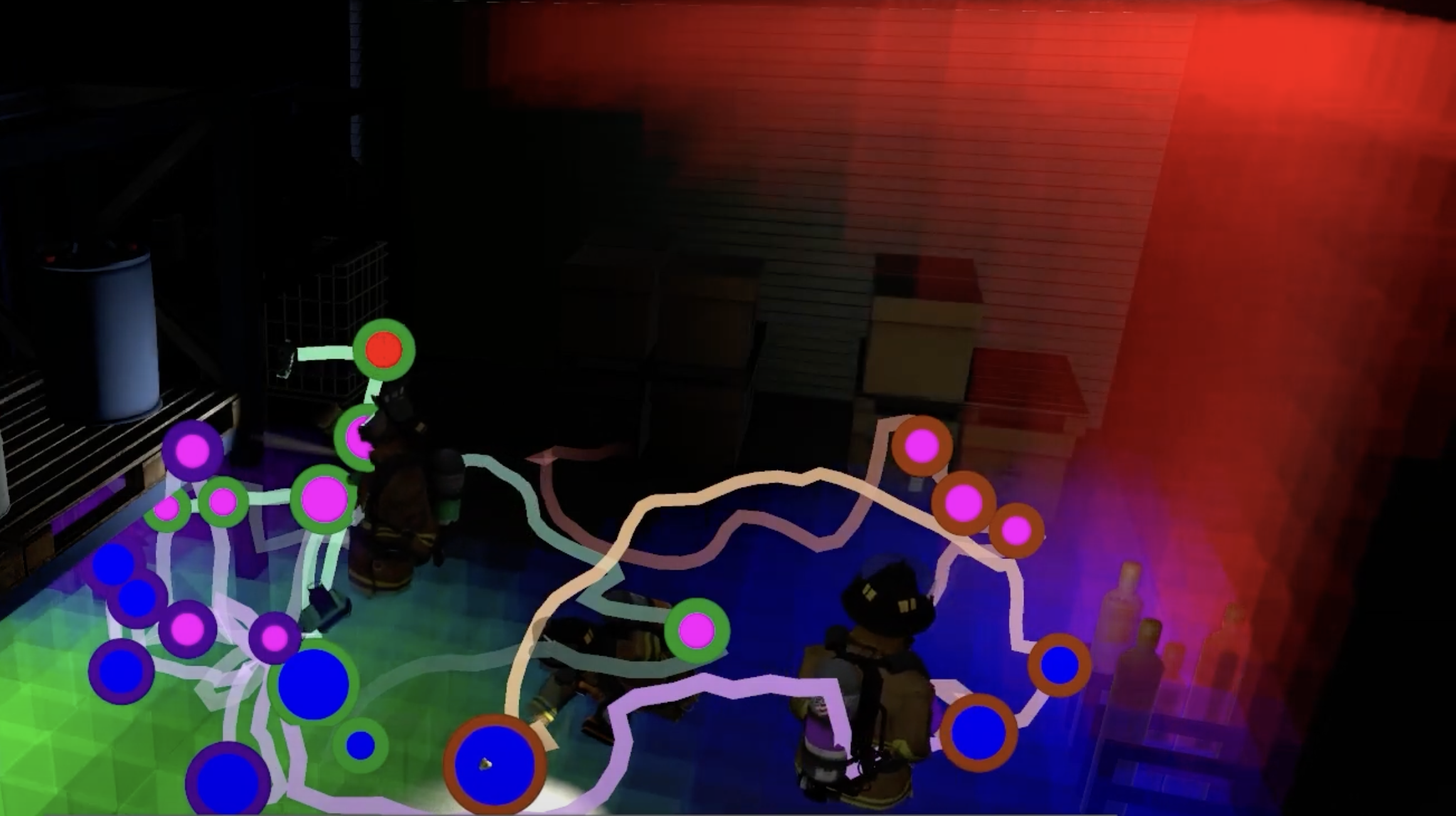}
 \caption{After action review of a gas monitoring training session, with a visualization of invisible gases (red and green volumes), along with the monitor's path, monitored locations, and thoroughness of the readings. Screengrab courtesy of NextGen Interactions~\cite{nextgen:2022aa}.}
 \label{fig:after-action}
\end{figure}

 In learning theory, error exposure training involves the use of someone's mistakes during training to provide performance feedback to the learner as part of the educational experience~\cite{rogers2002role}. In fact, learning from errors has been shown to be more effective than learning from one's successes~\cite{joung2006using}. In the case of gas monitoring tasks, where invisible but deadly gases need to be detected, if a HAZMAT technician in training fails to monitor certain areas for a sufficient amount of time, they have committed an error. A VR gas monitoring training tool can offer error exposure training by visually showing the location of invisible gasses that were missed by the trainee. Figure \ref{fig:after-action} is an example of a commercially available gas-monitoring training tool's after-action review system showing the location and concentrations of different gases and reading thoroughness.

\subsection{Physical props and passive haptics}
Typical VR equipment consists of a head-mounted display (HMD) and two controllers, which represent the user's head and hands. The controllers are capable of tracking the position and orientation of the user's hand, however, they have a fixed form factor. The user is able to interact with virtual objects and devices with their controllers, represented in VR by virtual hands, but they cannot feel the form and weight of these props. The nonexistent feeling of touch reduces immersion, can cause breaks in presence and limits the experience to the audio and visual channels only. This issue exists as long as standard VR controllers are used. \emph{Physical props} in VR are real tracked objects that are registered to digital twin counterparts displayed through the HMD~\cite{gainer2020customized}. As long as the user's hands can be tracked, the user can interact with the physical props by visualizing the prop's identical digital representation in the virtual environment along with the tactile feeling of holding and manipulating the physical object. The user is able to feel the texture, form, weight, and heat of these props, which addresses the limitations of standard VR controllers mapped to virtual objects.

Physical props are a type of \emph{passive haptics}~\cite{lindeman1999hand}. Passive haptics offer great realism to VR experiences by adding accurate somatosensory feedback to the user. Additionally to smaller props to represent handheld devices, passive haptics are also useful to represent larger structures, such as walls and furniture in virtual environments. Previous research demonstrated that using passive haptics in VR training applications produces a positive training outcome~\cite{insko2001passive}.

\subsection{Time perception in virtual reality}

The estimation of time is a subjective judgment that relates to perception, attention, and memory~\cite{matthews2016temporal}. Previous research found that VR causes a time compression effect, where users perceive time as running faster when in a VR simulation~\cite{schneider2011effect, mullen2021time}. However, there is some evidence that suggests no difference between perceived time in the real world and VR~\cite{bruder2014time}, particularly during short time estimates in the order of 2-5 seconds.

We hypothesize that differences in the perception of time between VR and the real world may be related to the realism of the experience. We suggest that increasing the physical realism of VR to approach that of the real world could lead to a better matching of the perceived time in VR to that of the real counterpart task.

Because gas monitoring tasks require precise time estimation, it is important that VR training for such ability approach the real world with respect to time estimation as much as possible. Thus, we investigated whether a VR training environment with high physical realism can cause time judgments to be closer to those in the real world.

\section{Testing Environment}

\begin{figure}[htb]
 \centering%
 \includegraphics[width=\columnwidth]{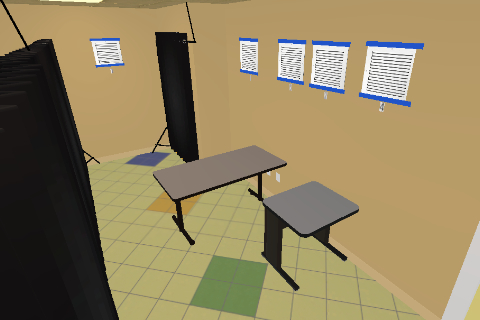}
 \caption{An overview of the virtual task environment. In the high physical realism condition, real desks matching the position of the virtual ones were placed in the experiment laboratory. In the standard VR condition (low physical realism), all real props were removed. The real-world condition was performed without the HMD and had all elements of the high physical realism VR condition. }
 \label{fig:overview}
\end{figure}

The experiment was conducted in a university lab with a section curtained off during the period in which the study was executed. Figure \ref{fig:overview} shows an overview of the virtual environment built for the study. The application was written in Unity and consisted of a hallway with a door through which the participant had to walk to take five gas leak measurements before returning to the origin (see Section \ref{sec:proc}).

The testing environment used in the experiment consisted of three setups, which were used as the experimental conditions. The conditions were the real world, which served as a baseline control, VR with high physical realism, and VR with low physical realism.

\textbf{\emph{Real world.}} In this setup, the user removed their HMD and completed the task with all the physical props. The gas monitoring device tracking was still used to indicate the start and end of holding the device up to the vent (see Section \ref{sec:proc}).

\textbf{\emph{VR with high physical realism.}} In the VR with high physical realism setup, there was a physical copy of all virtual environment objects, providing passive haptics. As Figure \ref{fig:RealMultiRAE} shows, the user picks up a real gas monitor, pushes a real door, and walks around real desks. 

\textbf{\emph{VR with low physical realism.}} In this setup, all physical props were removed, and the user interacted with the VR controller to pick up and hold a virtual gas monitor.

\section{Experiment}

Aiming at analyzing the effect that VR physical realism has on time perception, we conducted a repeated-measures experiment with 16 volunteers, all of whom were university students. The goal of the experiment was to evaluate how each VR condition (high physical realism vs. standard VR) compared to time estimation in the real world.

\subsection{Location}
The experiment was held in an isolated laboratory with minimized distractions. We created an artificial corridor by placing a wall of curtains parallel to a wall (see Figure \ref{fig:RealMultiRAE}A, B). This wall of curtain blocked visual distractions from the participant and reduced the size of the area we had to replicate in the virtual environment, leaving less room for errors.

\subsection{Participants}
A total of 16 individuals took part in this study, recruited through convenience sampling. Two participants did not provide their socio-demographic information. Of the 14 participants who reported socio-demographic information, 8 ($57\%$) were male and 6 ($43\%$) were female, with an average age of $M=20 \pm2.83$ years. When questioned about their familiarity with VR equipment, 6 ($43\%$) stated they were slightly familiar, 3 ($21\%$) moderately familiar, and 3 ($21\%$) not familiar at all. 8 individuals disclosed using corrective lenses for vision. 10 participants identified as right-handed, 3 as left-handed, and 1 as ambidextrous. When questioned about symptoms of dizziness, 7 ($50\%$) said "sometimes", 6 ($43\%$) said "never", and 1 ($7\%$) said "about half the time."

\subsection{Apparatus}
\begin{figure}[htb]
 \centering%
 \includegraphics[width=\columnwidth]{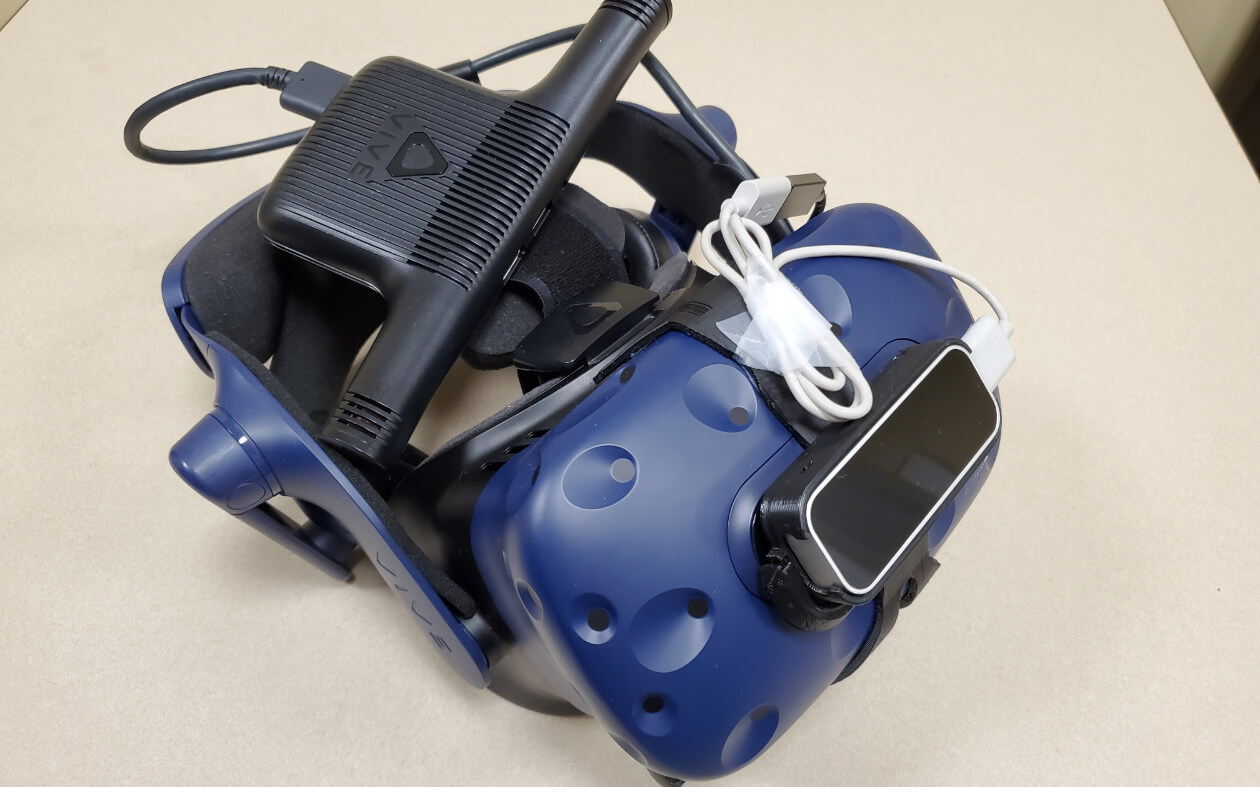}
 \caption{The HMD that was used in the study with the wireless transmitter and the Leap Motion hand tracker.}
 \label{fig:vive}
\end{figure}

The experiment was conducted over a Unity application. The hardware was an HTC Vive Pro with a wireless transmitter, and HTC Vive trackers (pucks) to track the gas monitoring device and the door in the high physical realism conditions. For the VR with high physical realism condition a Leap Motion hand tracker was mounted on the HTC Vive Pro (Figure \ref{fig:vive}). HTC Vive controllers provided tracking for the hands during the standard VR condition performance.

The physical props that provided passive haptics in the high physicality VR (and real world) condition were a 3D printed gas monitoring tool outfitted with inner weights to match the weight and center of gravity of a real device, a functional standalone door, and one small and one large desk for participants to maneuver between vents. Additionally, 5 vent images that matched the VR vent textures were printed in durable material.

\subsection{Design and data collection}
Each participant repeated the same set of tasks 3 times. All participants started with the real world condition to collect baseline time estimations, but the order for the VR conditions was counterbalanced to reduce potential order effects. Thus, the first condition was always the real world. Then participants performed either the standard VR condition or the high physical realism VR condition. Finally, the remaining condition was performed. Here, the independent variable was the condition performed by the participant.

Our goal was to make the participant interact with the props around them and get a feeling of the current level of physical realism, the participant would then go through a series of time estimation tests while we collected data. 

\subsection{Procedure} \label{sec:proc}
\begin{figure}[htb]
 \centering%
 \includegraphics[width=\columnwidth]{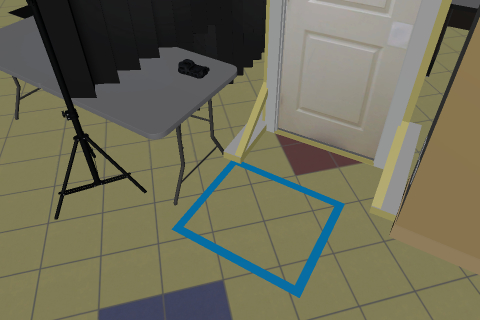}
 \caption{The participants started all tasks standing inside the blue square (which was also marked in the real world) and facing the door.}
 \label{fig:square}
\end{figure}

The participant walked into a small designated area marked on the ground (Figure \ref{fig:square}). In this area, the participant is provided the necessary equipment for that condition, such as the VR headset, controllers, and other tracking devices. If the condition is physical, a hand tracking system is activated to display their hands through the VR headset. The participant was reminded of the procedure before performing each condition. Participants were also told they would be asked to guess the total time they spent during the entire condition.

After each condition, the next one was prepared by having the experimenter facilitate the swapping of physical props for virtual ones and vice versa. For better tracking of the VR conditions, the black curtains were drawn after the participant was wearing the HMD, so they would not notice this physical environment change.

\begin{figure}[htb]
 \centering%
 \includegraphics[width=\columnwidth]{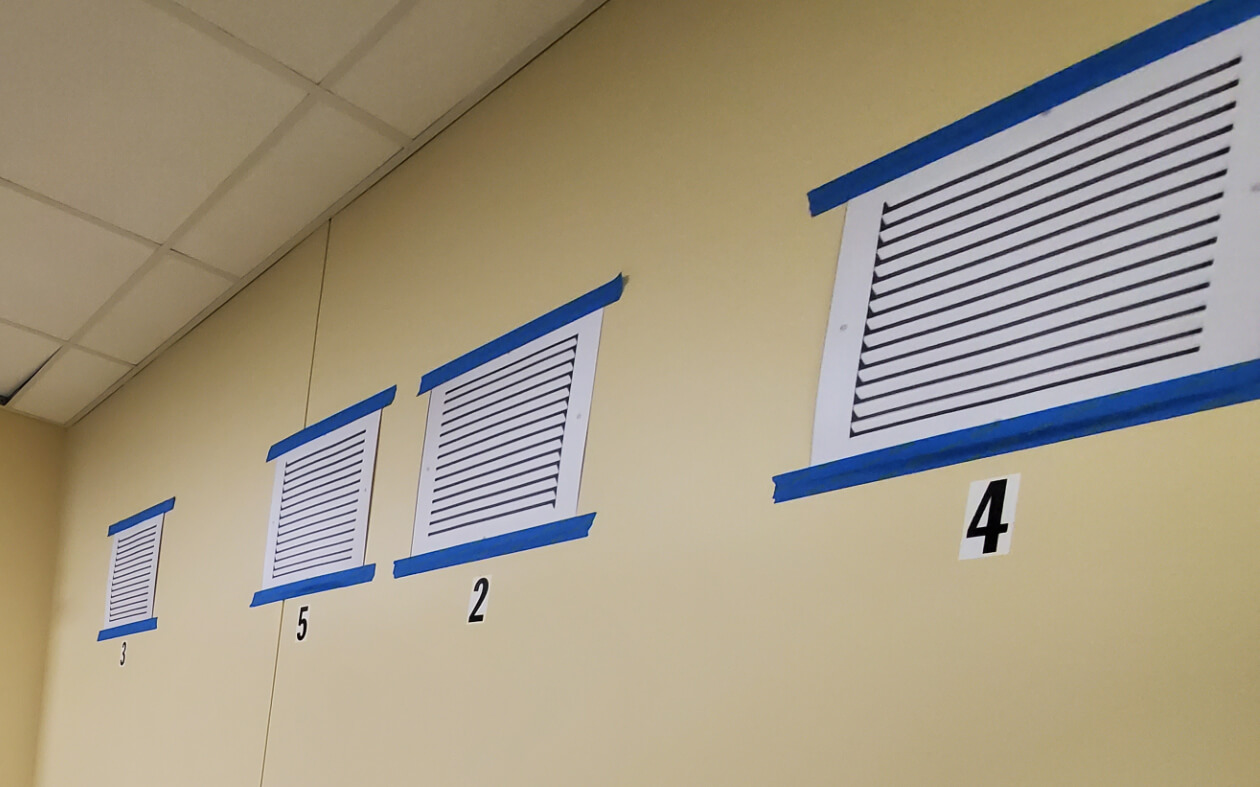}
 \caption{The order of vents to check from 2 to 5. Vent 1 can be seen in Figure 1B.}
 \label{fig:vents}
\end{figure}

With the experimenter's signal, the participant began by picking up the gas monitoring tool from the desk next to them (with or without physical props depending on the condition). The participant then opened the standalone door and walked through it. The participant walked up to the vent labeled "1" and brought the gas monitoring tool close to it. A signal was sounded at which time they were instructed to begin counting/estimating, in whole seconds, the time that passed until a second signal was heard. Unbeknownst to the participant, the duration between the two signals varied between 25 and 35 seconds, randomly selected  by the computer. After the second signal sounded, the participant verbally told the experimenter their estimation in seconds, and the researcher recorded that number. The participant repeated the time estimation process 4 more times in the order of the label of the vents (see Figure \ref{fig:vents}). After the participant finished the time estimations, they return back to the starting position and verbally told the researcher their rough estimate of the total time they spent during the condition. The computer recorded the actual time that passed between each of the 5 pairs of signals, and the total time the participant spent during each condition trial, starting upon crossing the door at the beginning and ending upon crossing the door back.

After each condition, the participant sat down and rested for 5 minutes.

\subsection{Data Analysis}
Data were collected during the experiment in terms of the participants' time estimation accuracy, and after the experiment as demographic data. From each participant, we collected 6 data points (5 vents and 1 total time) in each of the 3 conditions, totaling 240 data points that represent the participants' time estimation accuracy.

In this study, participants prospectively estimated the duration of a task limited by time. As such, each estimation task lasted a variable amount of time (see section \ref{sec:proc}), so we had to normalize the estimates to make estimates comparable. Thus, the main dependent variable for time estimation is \textit{accuracy}, which is measured as the ratio between the estimated time and the observed time:
\vspace{-3pt}
\[ accuracy = \frac{estimatedTime}{elapsedTime}.\]
\noindent This way, the closer to $1$ $accuracy$ is, the more accurate the estimate. Ratios $<1$ indicate time compression and ratios $>1$ indicate time dilation.

\section{Results}

Recall that participants had nested time estimation tasks. A short period counting task that simulates gas leak monitoring (repeated over 5 vents) and the estimate of the entire trial period for the interface condition. We present the results for each of these two task levels.

Table \ref{tab:sum} shows the means and standard deviations of the accuracy over condition for the vent time estimation tasks and for the total trial time estimation tasks.

\begin{table}
\caption{Mean and SD accuracy for time estimates per condition.}
\label{tab:sum}
  \centering
  \begin{tabular}{lrrr}
    \toprule
    Condition & Estimate & Mean Accuracy & SD Accuracy \\ 
     \midrule
    Real & Vent 1 & 1.130 & 0.410 \\ 
    Real & Vent 2 & 1.050 & 0.292 \\ 
    Real & Vent 3 & 1.094 & 0.361 \\ 
    Real & Vent 4 & 1.030 & 0.273 \\ 
    Real & Vent 5 & 1.101 & 0.350 \\ 
    Real & Total Time & 1.421 & 0.763 \\ 
    \midrule
    High & Vent 1 & 1.099 & 0.378 \\ 
    High & Vent 2 & 1.097 & 0.474 \\ 
    High & Vent 3 & 1.040 & 0.369 \\ 
    High & Vent 4 & 1.004 & 0.320 \\ 
    High & Vent 5 & 1.022 & 0.293 \\ 
    High & Total Time & 1.432 & 0.751 \\ 
     \midrule
    Low & Vent 1 & 1.062 & 0.302 \\ 
    Low & Vent 2 & 1.155 & 0.627 \\ 
    Low & Vent 3 & 1.088 & 0.368 \\ 
    Low & Vent 4 & 1.042 & 0.271 \\ 
    Low & Vent 5 & 1.028 & 0.299 \\ 
    Low & Total Time & 1.378 & 0.732 \\ 
     \bottomrule
  \end{tabular}
  \end{table}  
  
\subsection{Gas Leak Monitoring Time Estimation}

\begin{figure}[htb]
 \centering
 \includegraphics[width=\columnwidth]{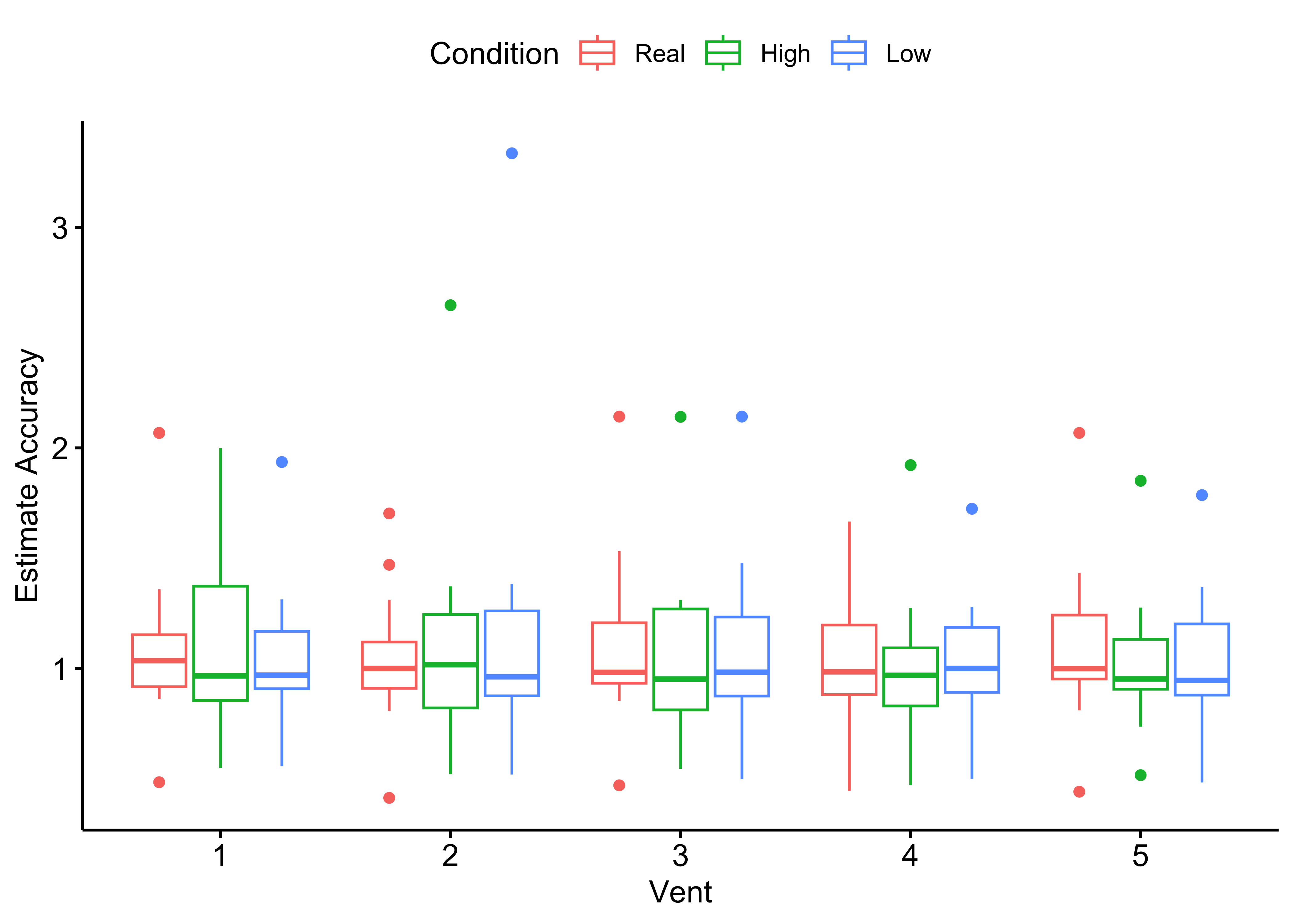}
 \caption{Overall vent time estimation results for accuracy. No main or interaction effects were statistically significant.}
 \label{fig:accuracy}
\end{figure}

Figure \ref{fig:accuracy} shows the overall results for the gas leak monitoring task. The mean \textit{accuracy} for participants when considering only condition (real world, high physical realism VR, and standard VR) was $M=1.08 \pm0.33$, $M = 1.05 \pm0.36$, and $M = 1.07\pm0.38$, respectively.

A repeated measures ANOVA was conducted to investigate the effect of Condition and Vent on Accuracy. The results show that the main effect of the Condition was not statistically significant, $F(2, 28) = 0.594$, $p = 0.559$. This suggests that there was no significant difference in the accuracy between the different condition groups. The main effect of Vent was also not statistically significant, $F(1,98, 27.69) = 1.219$, $p = 0.311$. This suggests that there was no significant difference in the accuracy between the different Vent groups. The interaction effect of Condition and Vent was not statistically significant, $F(2,22, 31.12) = 1.472$, $p = 0.245$. This indicates that there was no significant interaction between the Condition and Vent groups on Accuracy. Because there were no main or interaction effects on Vent, it is reasonable to aggregate the readings for the 5 vents (at which the time estimation task was identical) for each participant. Thus, the remainder of the data analysis was performed on the aggregated vent estimates for each participant.

\begin{figure}[htb]
 \centering
 \includegraphics[width=\columnwidth]{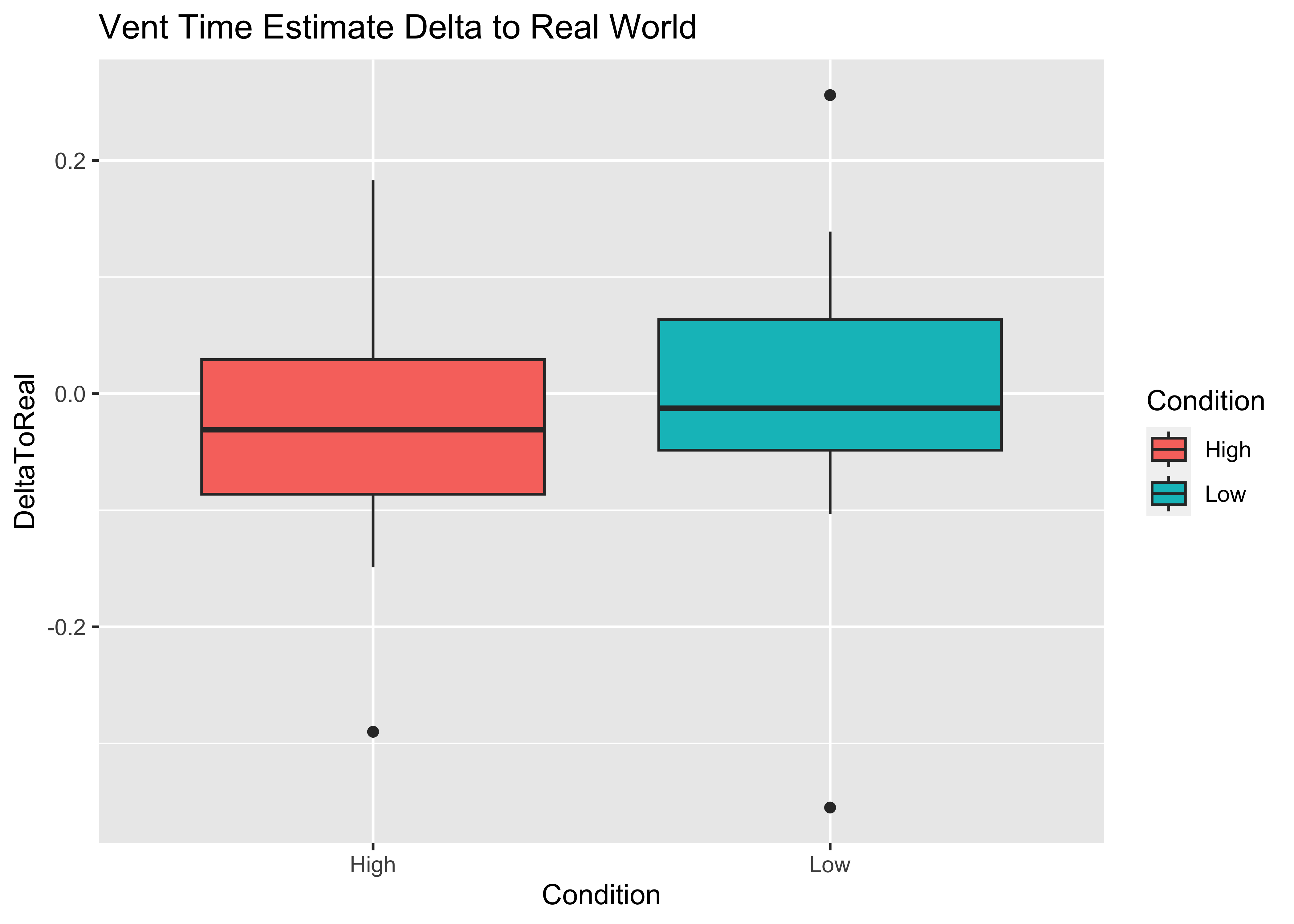}
 \caption{Delta difference between the experimental conditions accuracy and the real world accuracy.}
 \label{fig:ventdeltas}
\end{figure}

As a training environment, it is important to understand the effect that the training interface may have over the environment in which the trained task is executed--the real world. So, we computed the difference (delta) in accuracy between the \textit{experimental} VR condition (high or low physical realism) and the real world \textit{control} condition. Figure \ref{fig:ventdeltas} shows the difference between the deltas for the high and low experimental conditions. A paired t-test comparing the deltas of each condition to control showed no statistical significance, $t(15)=-1.061$, $p=0.305$. This suggests that there was no significant difference across the accuracy deltas of each experimental condition to the real world.

To further investigate the similarity between the accuracy in the experimental conditions and the real world, we ran correlation analyses. There was a strong positive correlation between the accuracy in the Low physical realism VR condition and the Real World, $r(14)=.929, p<.0001$. Likewise, the High and Real conditions were also strongly correlated,  $r(14)=.946, p<.0001$. Finally, there was a strong positive correlation between the accuracy in the Low and High conditions,  $r(14)=.964, p<.0001$. These correlations indicate that the interface condition (high physical realism VR, low physical realism VR, real world) may not have influenced time estimation.

\subsection{Total Trial Time Estimation}

\begin{figure}[htb]
 \centering
 \includegraphics[width=\columnwidth]{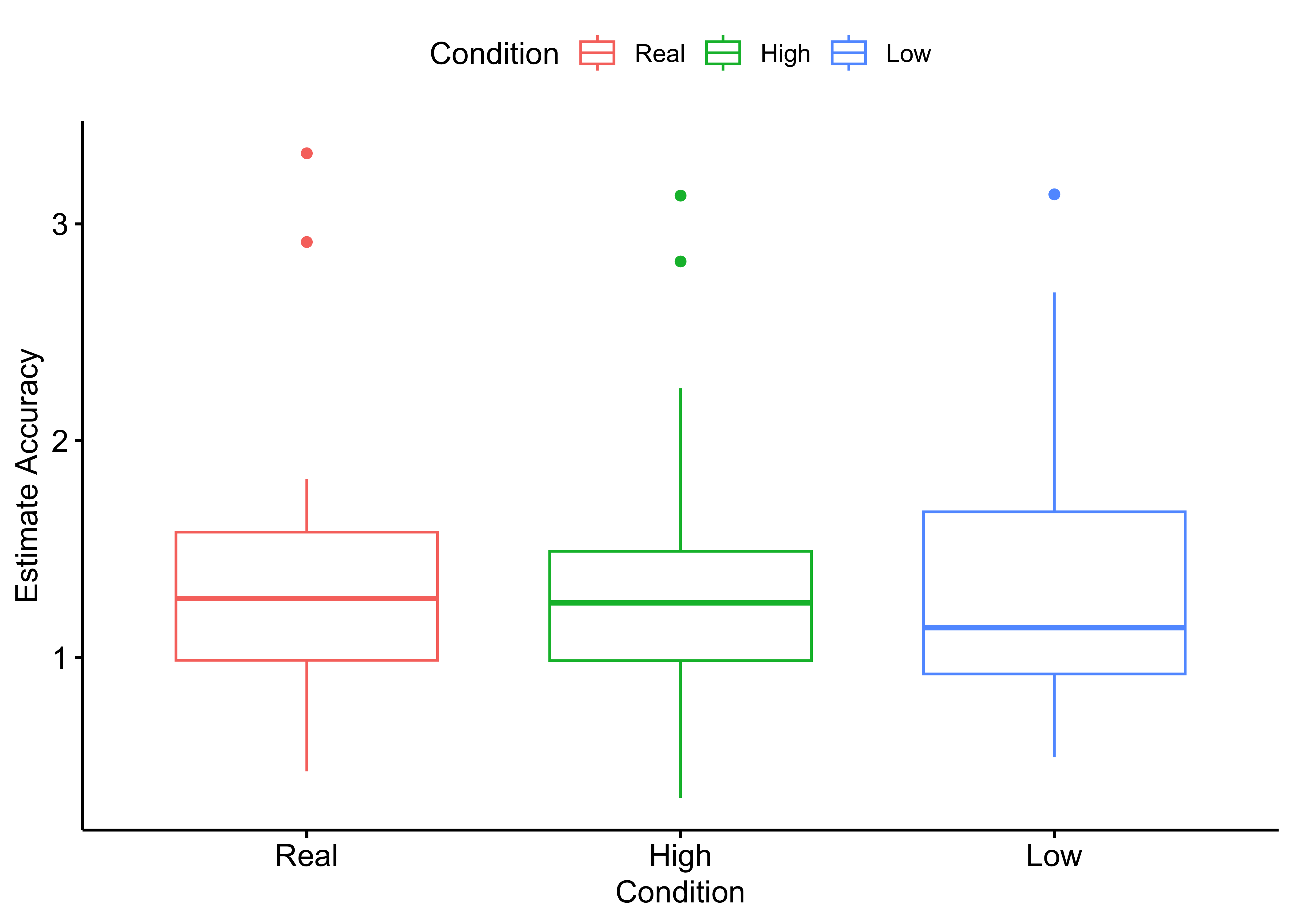}
\caption{Overall total time estimation results for accuracy. No main effects were statistically significant.}
  \label{fig:total}
\end{figure}

Figure \ref{fig:total} shows the results of the total trial time estimation. The mean accuracy for participants when considering only the real world conditions was $M = 1.42 \pm0.76$, for high physical realism VR it was $M = 1.43 \pm0.75$, and for standard VR, it was $M = 1.37 \pm0.73$. In the repeated measures ANOVA, the results found that the main effect of the Condition was not statistically significant, with $F(1.32, 18.45) = 0.003$, $p = 0.981$. This suggests that there was no significant difference in the accuracy between the different condition groups.

\begin{figure}[htb]
 \centering
 \includegraphics[width=\columnwidth]{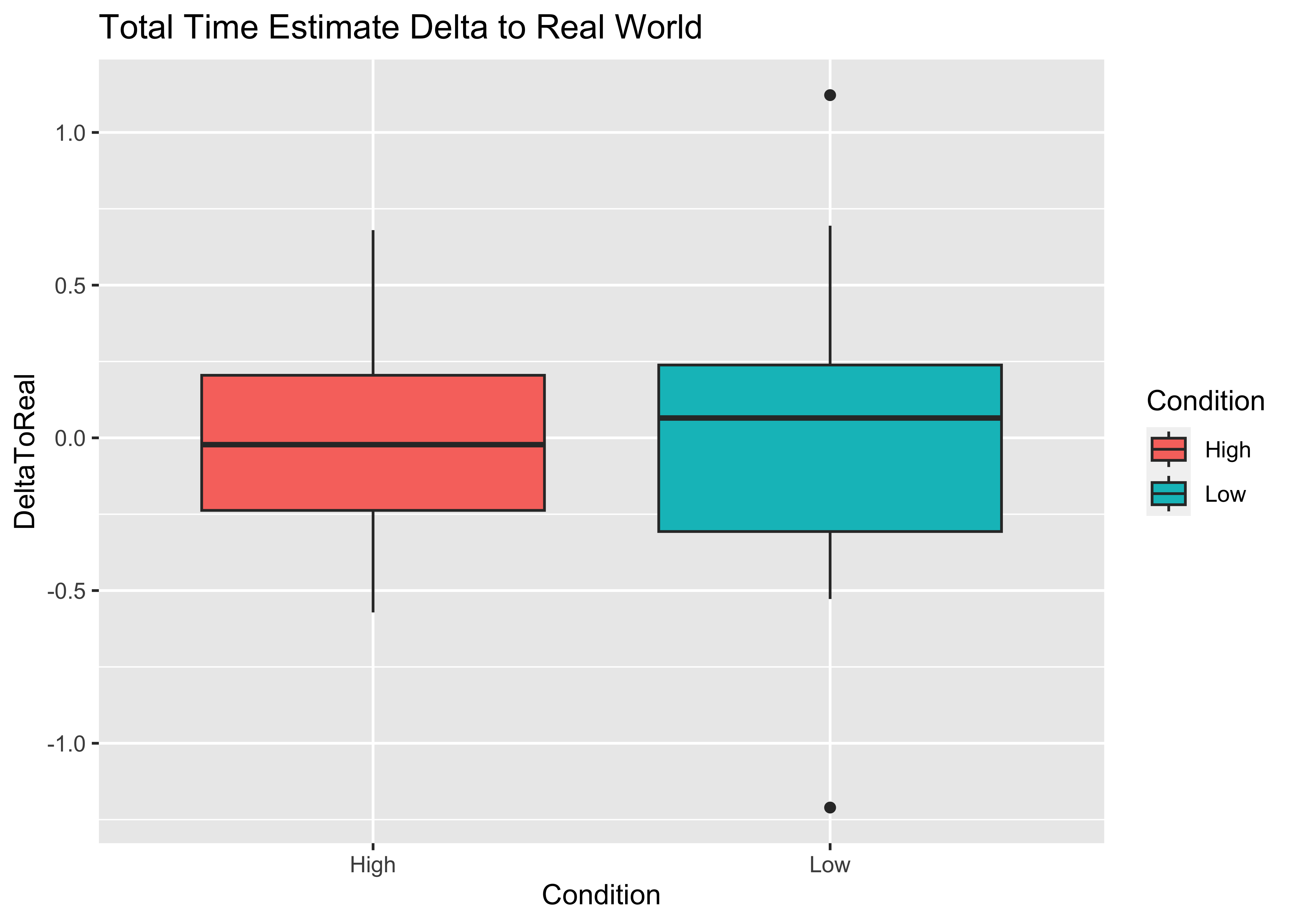}
\caption{Overall total time estimation results for accuracy. No main effects were statistically significant.}
  \label{fig:deltatotal}
\end{figure}

We also looked at the delta between the real world and each condition estimate for the total trial time (Figure \ref{fig:deltatotal}). No statistical significance was found in the paired t-test, $t(13)=.026, p=.980$. As with the vent estimates, the deltas are very close to zero, which gives evidence that time estimation was not different between the experimental conditions and the real world. This result is also corroborated by strong positive correlations between all conditions.

\section{Discussion}

\subsection{Time Dilation}

To our surprise, the results consistently showed a \textit{time dilation} effect as opposed to an expected time compression effect, regardless of the environment that was tested. In most time estimation studies from the literature, participants are instructed to estimate time until they feel that a fixed duration, such as 5 minutes, has passed~\cite{mullen2021time}. We are intrigued by this seeming contradicting results. We speculate that this may be partially explained by the nature of the gas leak monitoring task. During those tasks, participants had to hold the gas monitor up to the vent, which may have caused fatigue leading to a perception of the time passing more slowly. Each estimated second would be counted faster due to an anxiety response caused by fatigue.

In hindsight, the findings are consistent with what hazmat instructors have told us: they get frustrated that hazmat personnel rarely hold the air monitor for sufficient time to get accurate readings. One potential explanation is that holding a device in a single position for 30 seconds is a boring task (as well as perhaps a fatiguing task for a heavy monitor). Similarly to the old saying that a watched pot never boils, time can seem to pass slower when waiting for something to happen.

However, this reasoning does not seem to fully explain the time dilation effect. First, time dilation, although present in the gas leak monitoring task, was rather mild, with accuracy ratios very close to one. In fact, looking at the frequencies of time dilation vs. time compression estimates in the gas leak monitoring tasks, participants reported dilated estimates in 101 trials overall and compressed estimates in 131 trials. This means that time compression was more frequent than time dilation for the short task, but dilated estimates typically had higher magnitude than compressed estimates. By contrast, the long total time estimates showed much more consistent time dilation, with 34 dilated estimates and 13 compressed estimates.

Given that time dilation still happened, and was stronger, in the total task time estimation, we don't believe that fatigue alone should be attributed for the counter intuitive time dilation effect. We speculate that, when participants are not in control of the duration that they remain in the experience, there are different time estimation mechanisms at play than those when participants are responsible to guess when time is up. Future work comparing proactive and retroactive time estimation in VR is needed to verify this hypothesis.

\subsection{Time Estimations Across Conditions}

Our study found that, although we did not observe significant effects, time dilation was consistently present throughout the trials. The fact that similar time dilation was found in the real-world and both virtual reality conditions suggests that time dilation occurs regardless of the environment in which the task is performed for tasks such as counting time between signals and retroactively estimating time spent in a longer task.

An interesting observation is that the time dilation for the total task time estimation ($\sim1.4$) was significantly higher than the time dilation observed in the short vent tasks ($\sim1.07$). We cannot say with certainty the reason for such a phenomenon, but we speculate that focused attention may have contributed to more accurate estimations in the vent task.

To the best of our knowledge, this study was the first to compare time estimation in different virtual reality conditions using a real-world estimate as baseline. We did find that time dilation occurred in all conditions, which suggests that the task itself may have had the strongest impact on time estimation, rather than the environment in which the task was performed. It's possible that humans naturally tend to overestimate time when performing conscious, methodical tasks such as counting time between two signals, and that any potential effects of the interface condition may have been obscured by the task's impact on time estimation.

Another possible explanation for the lack of observed effects is the use of a within-subject experimental design. Previous research has found a significant time estimation bias during repeated exposures, which suggests that between-subject experiments may be more appropriate for studying differences in time estimation between conditions~\cite{mullen2021time}. Further research, involving more time and participants, is needed to verify this assumption.

\acknowledgments{This work was performed under award \#70NANB18H146 from
the U.S. Department of Commerce, National Institute of Standards
and Technology, Public Safety Communications Research Division.
We thank all firefighters who provided input and feedback on this project.}

\balance

\bibliographystyle{abbrvurl}

\bibliography{template}
\end{document}